\documentclass[12pt]{article}

\usepackage{amsmath, amsthm, amsfonts, amssymb}

\def\P{\mathbb{P}}
\def\E{\mathbb{E}}

\newcommand{\rme}{{\rm e}}
\newcommand{\rmd}{{\rm d}}

\newcommand{\1}{{\bf 1}}

\def\ep{\varepsilon}
\def\epi{\varepsilon_i}

\newtheorem{theo}{Theorem}[section]

\newtheorem{rem}{Remark}[section]

\newtheorem{cor}{Corollary}[section]

\def\call{{\rm c\kern-3.5pt                                    
 \vrule height 5.0pt width 0.4pt depth -0.5pt \phantom {.}}}
\makeatother

\makeatletter \@addtoreset{equation}{section}

\def\theremark{\arabic{section}.\arabic{remark}\protect\@blinkpoint}
\def\@blinkpoint{.}
\let\@blinkref=\ref
\def\ref#1{{\def\@blinkpoint{}\@blinkref{#1}}}
\makeatother

\begin{document}

\title{Option Pricing Model Based on a Markov-modulated Diffusion with Jumps} 

  \author{Nikita Ratanov\footnote{ Universidad del Rosario, Bogot\'a,
Colombia}}
\date{}
\maketitle

\begin{abstract}
The paper proposes a class of financial market models which are
based on  inhomogeneous telegraph processes and jump diffusions with alternating volatilities.
It is assumed that the jumps
occur when the tendencies and volatilities are switching.
We argue that such a model captures well the stock price dynamics under  periodic financial cycles.
The distribution of this process is described in detail. For this model we obtain the
structure of the set of martingale measures. This incomplete model can be
completed by adding another asset based on the same sources of
randomness. Explicit closed-form formulae for prices of the
standard European options are obtained for the completed market model.
\end{abstract}


\section{Introduction}

Beginning with the works of Mandelbrot (1963), Mandelbrot and Taylor (1967),
 and Clark (1973), it is commonly accepted that
 the dynamics of asset returns cannot be described by
 geometric Brownian motion with constant parameters of drift and volatility.
 A lot of sophisticated constructions have been exploited
 to capture the features
 that help to express the reality better than Black-Scholes-Merton model.
 Merton (1976) which have incorporated jump diffusion model for the asset price was the first.
Later on the constructions with random
 drift and random volatility parameters appeared.
 A popular approach is to use L\'evy processes
 with stationary independent increments. However, this theoretical behavior
 does not match empirical observations.

Another approach utilizes markovian dependence on the past and the technique
of Markov random processes (see Elliott and van der Hoek (1997)).
We deal mainly with this direction. More precisely, the model is based
on a standard Brownian motion $w=w(t), t\geq 0$ and on
a Markov process $\ep(t), t\geq 0$ with two states $0, 1$ and transition
probability intensities $\lambda_0$ and $\lambda_1$.

Let us define processes $c_{\ep(t)}$, $\sigma_{\ep(t)}$ and $r_{\ep(t)}, t\geq 0$, where $c_0\geq c_1, r_{0}, r_1>0$. Then, we introduce $\mathcal{T}(t)=\int_0^tc_{\ep(\tau)}\rmd\tau$,
$\mathcal{D}(t)=\int_0^t\sigma_{\ep(\tau)}\rmd w(\tau)$ and a pure
jump process $\mathcal{J}=\mathcal{J}(t)$ with alternating jumps
of sizes $h_0$ and $h_1$, $h_0, h_1>-1$.

The continuous time random motion $\mathcal{T}(t)=\int_0^tc_{\ep(\tau)}\rmd\tau, t\geq 0$
with alternating velocities is known as telegraph process. This type
of processes have been used before in various probabilistic aspects (see, for instance, Goldstein (1951),
Kac (1974) and Zacks (2004)). These processes have been
exploited for stochastic volatility modeling (Di Masi et al
(1994)), as well as for obtaining a ``telegraph analog" of the
Black-Scholes model (Di Crescenzo and Pellerey (2002)).
The option pricing models based on continuous-time random walks are
widely presented in the physics literature (see Masoliver et al (2006) or Montero (2008)).
Recently the telegraph processes was applied to actuarial problems,
Mazza and Rulli\`{e}re (2004). Markov-modulated diffusion process
$\mathcal{D}(t)=\int_0^t\sigma_{\ep(\tau)}\rmd w(\tau)$ was exploited for financial market modeling
(see Guo (2001), Jobert and Rogers (2006)), as well as in insurance (see B\"{a}uerle and K\"{o}tter (2007))
or in theory of queueing networks (see Ren and  Kobayashi (1998)).

This paper deals with the market model which presumes the
 evolution of risky asset $S(t)$ is given
by the stochastic exponential of the sum $\mathcal{X}=\mathcal{T}(t)+\mathcal{D}(t)+\mathcal{J}(t)$.
The bond price is the usual exponential of the process
$\mathcal{Y}=\mathcal{Y}(t)=\int_0^tr_{\ep(\tau)}\rmd\tau, t\geq 0$ with alternating
interest rates $r_0$ and $r_1$.

This model generalizes classic  Black-Scholes-Merton model based on geometric Brownian motion
($c_0=c_1,\ r_0=r_1,\ \sigma_0=\sigma_1\neq 0,\ h_0=h_1=0$), Black and Scholes (1973), Merton (1973).
Other  particular versions of this model was also discussed before:
\begin{enumerate}
\item $c_0=c_1,\ \sigma_0=\sigma_1=0,\ h_0=h_1\neq 0$:  \textit{ Merton model}, Merton (1976), Cox and Ross (1976);
\item $c_0\neq c_1,\ \sigma_0=\sigma_1=0,\ h_0\neq h_1$:  \textit{jump-telegraph model}, Ratanov (2007);
\item $c_0\neq c_1,\ \sigma_0\neq\sigma_1,\ h_0=h_1=0$: \textit{Markov-modulated dynamics}, Guo  (2001), Jobert and Rogers (2006).
 \end{enumerate}

The jump-telegraph model, as well as Black-Scholes and Merton model,
is free of arbitrage opportunities, and it is complete.
Moreover it permits explicit standard option pricing formulae
similar to the classic Black-Scholes formula. Under suitable rescaling
this model converges to the Black-Scholes (see Ratanov (2007)).
First calibration results of the parameters of the telegraph model
have been presented in  De Gregorio and Iacus (2007).
These estimations have been based on the data of Dow-Jones industrial average (July 1971 - Aug 1974).
However, a presence of jumps and/or diffusion components has not been estimated.
Nevertheless, an implied volatility with respect to a moneyness variable in stochastic
volatility models of the Ornstein-Uhlenbeck type (see Nicolato and Venardos (2003)) looks very similar to
the volatility smile in jump telegraph model (see Ratanov (2007b)).

In this paper we extend the jump-telegraph market model, presented in
Ratanov (2007), by adding the diffusion component
with alternating volatility coefficient.

The jump-telegraph model equipped with the diffusion term becomes more realistic.
Indeed, the alternating velocities of the telegraph process describe long-term financial trends,
 and the diffusion summand introduces an uncertainty of current prices.
 This uncertainty may has different volatilities in the bearish and in the bullish trends
 ($\sigma_0\neq\sigma_1$).

The paper  is organized as follows: in
Section \ref{sec1} we present the detailed definitions and the
description of underlying processes and their distributions.
The explicit construction of a measure change is given by
the Girsanov theorem for jump
telegraph-diffusion processes.

In Section \ref{secmod} we describe the set of risk-neutral
measures for the  incomplete jump telegraph-diffusion model. Also
we consider a completion of the model by adding another asset
driven by the same sources of randomness. For the completed market model
we obtain explicit option pricing formulae of the standard call
option. These formulae are based on a mix of Black-Scholes function and
densities of spending times of the driving Markov flow.

\section{Jump telegraph processes and jump diffusions with Markov
switching}\label{sec1}

Let $(\Omega, \mathfrak{F}, \P)$ be a probability space.
Denote $\epi(t),\ t\geq 0,\ i=0, 1$ a pair of Markov processes with two states
$\{0, 1\}$ and with rates $\lambda_0, \lambda_1>0$:

\[
\P\{\epi(t+\Delta t)=j\ |\ \epi(t)=j\}=1-\lambda_j\Delta t+o(\Delta
t),\quad \Delta t\to 0,\ i, j=0, 1.
\]
Subscript $i$ indicates the initial state: $\ep_i(0)=i$.

Let $\tau_1, \tau_2, \ldots$ be switching times.  The
time intervals $\tau_j-\tau_{j-1},\ j=1,\ 2,\ \ldots$ ($\tau_0=0$),
separated by moments of value changes $\tau_j=\tau_j^i$ are
independent and exponentially distributed. We denote by $\P_i$ the conditional probability with
respect to the initial state $i=0, 1$, and by $\E_i$ the expectation
with respect to $\P_i$.

 Denote by $N_i(t)=\max\{j: \tau_j\leq t\}, t\geq 0$ a number of switchings of
$\epi$ till time $t,\ t\geq 0$. It is clear that $N_i,\ i=0, 1$ are
the counting Poisson processes with alternating intensities
$\lambda_0, \lambda_1>0$. It is easy to see that the distributions $\pi_i(t;
n):=\P_i\{N_i(t)=n\}, n=0, 1, 2, \ldots,\ i=0,1,\ t\geq 0$ of the
processes $N_i=N_i(t)$ satisfy the following system:

\begin{eqnarray}\label{eqpi}
\frac{\rmd\pi_i(t; n)}{\rmd t}&=&-\lambda_i\pi_i(t;
n)+\lambda_i\pi_{1-i}(t; n-1),\quad i=0, 1,\ n\geq 1,\\ \nonumber
\pi_i(t; 0)&=&\rme^{-\lambda_it}.
\end{eqnarray}

To prove it notice that conditioning on the Poisson event on the
time interval $(0, \Delta t)$ one can obtain
\[
\pi_i(t+\Delta t; n)=(1-\lambda_i\Delta t)\pi_i(t;
n)+\lambda_i\Delta t\pi_{1-i}(t; n-1)+o(\Delta t),\quad \Delta t\to
0,
\]

\noindent which immediately leads to (\ref{eqpi}).

Let $c_0, c_1,\ c_0>c_1;\  h_{0}, h_{1}; \sigma_{0}, \sigma_{1}$
be real numbers. Let $w=w(t),\ t\geq 0$ be a standard Brownian
motion independent of $\epi$. We consider

\begin{eqnarray}\label{tjd}
\mathcal{T}_i(t)=\mathcal{T}_i(t; c_0,
c_1)=\int\limits_0^tc_{\ep_i(\tau)}\rmd\tau, &\qquad
\mathcal{J}_i(t)=\mathcal{J}_i(t; h_0,
h_1)=\int\limits_0^th_{\ep_i(\tau)}\rmd
N_i(\tau)=\sum\limits_{j=1}^{N_i(t)}h_{\epi(\tau_j-)},\nonumber
\\
&\mathcal{D}_i(t)=\mathcal{D}_i(t; \sigma_0,
\sigma_1)=\int\limits_0^t\sigma_{\ep_i(\tau)}\rmd w(\tau).&
\end{eqnarray}

Processes $\mathcal{T}_0, \mathcal{T}_1$ are 
telegraph processes with the states $<c_0, \lambda_0>$ and $<c_1,
\lambda_1>$, $\mathcal{J}_0, \mathcal{J}_1$ have a sense of pure
jump processes, and $\mathcal{D}_0, \mathcal{D}_1$ are Markov-modulated
diffusions. Thus the sum
$\mathcal{X}_i:=\mathcal{T}_i(t)+\mathcal{J}_i(t)+\mathcal{D}_i(t),\
t\geq 0,\ i=0, 1$ is naturally called jump telegraph-diffusion (JTD)
process with two
states, $<c_0, h_0, \sigma_0, \lambda_0>$ and $<c_1, h_1, \sigma_1, \lambda_1>$. 

Further, we will assume all processes to be adapted to the natural
filtration $\mathfrak{F}^i=(\mathfrak{F}_t^i)_{t\geq 0}$
($\mathfrak{F}_0^i=\{\emptyset,\ \Omega\}$), generated by
$\ep_i(t), \ t\geq 0$, and $w(t), \ t\geq 0$. We suppose that the
filtration satisfies the ``usual conditions" (see e. g.
Karatzas and Schreve (1998)).

The distribution of $\mathcal{X}_i(t)$
can be found exactly. First, we denote by $p_i(x, t; n)$
 (generalized) probability
densities with respect to the measure $\P_i$ of the jump telegraph-diffusion
variable $\mathcal{X}_i(t)$,
which has $n$ turns up to time $t$:

\begin{equation}\label{JTdensity}
\P_i\{\mathcal{X}_i(t)\in\Delta, \
N_i(t)=n\}=\int\limits_\Delta p_i(x, t; n)\rmd x, \quad i=0, 1,\
t\geq 0,\ n=0, 1, 2, \dots
\end{equation}

The PDEs which describe the densities $p_i(x,  t; n)$ have the following form.

\begin{theo}
Densities $p_i, i=0, 1$ satisfy the following PDE-system
\begin{equation}\label{eqDT}
\frac{\partial p_i}{\partial t}(x, t; n)+c_i\frac{\partial
p_i}{\partial x}(x, t; n)-\frac{\sigma_i^2}{2}\frac{\partial^2
p_i}{\partial x^2}(x, t; n)=-\lambda_ip_i(x, t; n)+\lambda_i
p_{1-i}(x-h_i, t; n-1),\ t>0,
\end{equation}
\[
\qquad i=0, 1,\ n\geq 1.
\]
Moreover
\begin{equation}\label{n0}
p_i(x, t; 0)=\rme^{-\lambda_it}\psi_i(x, t),
\end{equation}
where
\begin{equation}\label{psii}
\psi_i(x, t)=\frac{1}{\sigma_i\sqrt{2\pi
t}}\rme^{-\frac{(x-c_it)^2}{2\sigma_i^2t}},
\end{equation}
and
\[
p_i(x, t; n)|_{t\downarrow 0}=0,\ n\geq 1,\ i=0, 1.
\]
\end{theo}

\proof The equality (\ref{n0}) follows from definitions (\ref{tjd})-(\ref{JTdensity}).

To derive (\ref{eqDT}) note that from the properties of Poisson and Wiener processes (see e.g. Protter (1990))
for any $t_2>t_1$ it follows that
\begin{equation}\label{XX'}
\mathcal{X}_i(t_2)
=\mathcal{X}_i(t_1)
+\mathcal{X}'_{\epi(t_1)}(t_2-t_1),
\end{equation}
where $\mathcal{X}'_i$ is a copy of the process
$\mathcal{X}_i$, $i=0, 1$
 which is independent of the original.

Let $\Delta t>0$.
From (\ref{XX'}) it follows that
$\mathcal{X}_i(t+\Delta t)=\mathcal{X}_i(\Delta t)+\mathcal{X}'_i(t)$.
Let $\tau$ is the random variable uniformly
distributed on $[0,\ \Delta t]$ and independent of $\mathcal{X}_i$. Notice that 
$\mathcal{X}_i(\Delta t)=c_i\Delta t+\sigma_iw(\Delta t)$, if $N_i(\Delta t)=0$, and
$\mathcal{X}_i(\Delta t)\stackrel{d}{=}c_i\tau+c_{1-i}(\Delta t-\tau)+\sigma_iw(\tau)+\sigma_{1-i}w(\Delta t-\tau)+h_i$,
if $N_i(\Delta t)=1$.

Since $\P_i(N_i(\Delta t)>1)=o(\Delta t)$ as $\Delta t\to 0$, then conditioning on a jump in $(0, \Delta t)$ we have

\begin{equation}\label{eqpipre}
p_i(x, t+\Delta t; n)=(1-\lambda_i\Delta t)p_i(\cdot, t;
n)*\psi_i(\cdot, \Delta t)(x)+\lambda_i\Delta t p_{1-i}(\cdot, t;
n-1)*\tilde\psi_i(\cdot, \Delta t)(x-h_i)+o(\Delta t),
\end{equation}

\noindent $i=0, 1,\ \Delta t\to 0$. Here $\psi_i(\cdot, \Delta t)$, the
distribution density of $c_i\Delta t+\sigma_iw(\Delta t)$, is defined in (\ref{psii}), and
$\tilde\psi_i(\cdot, \Delta t)$ is the distribution density of
$c_i\tau+c_{1-i}(\Delta
t-\tau)+\sigma_iw(\tau)+\sigma_{1-i}w(\Delta t-\tau)$; the
notation  $*$ is used for the convolution in spacial variables.

It is easy to see, that $\psi_i(x, \Delta t),\ \tilde\psi_i(x,
\Delta t)\to \delta(x)$ as $\Delta t\to 0$. Hence

\begin{equation}\label{convpi}\nonumber
p_i(\cdot, t; n)*\psi_i(\cdot, \Delta t)(x)\to p_i(x, t; n),
\end{equation}

\begin{equation}\label{convpi1}
p_{1-i}(\cdot, t;
n-1)*\tilde\psi_i(\cdot, \Delta t)(x-h_i)\to p_{1-i}(x-h_i, t; n-1)
\end{equation}

\noindent as $\Delta t\to 0$.

 Then,

\[
\frac{1}{\Delta t}\left[p_i(\cdot, t; n)*\psi_i(\cdot, \Delta
t)(x)-p_i(x, t; n)\right]=\frac{1}{\Delta
t}\left[\int\limits_{-\infty}^\infty p_i(x-y, t; n)\psi_i(y,
\Delta t)\rmd y-p_i(x, t; n)\right]
\]
\[
=\frac{1}{\Delta t}\int\limits_{-\infty}^\infty
\left[p_i(x-c_i\Delta t-y\sigma_i\sqrt{\Delta t}, t; n)-p_i(x, t;
n)\right]\psi(y)\rmd y,
\]

\noindent where $\psi=\psi(\cdot)$ is $\mathcal{N}(0, 1)$-density.
The latter value equals to

\[
\frac{1}{\Delta
t}\int\limits_{-\infty}^\infty\psi(y)\left[\frac{\partial
p_i}{\partial x}(x, t; n)(-c_i\Delta t-y\sigma_i\sqrt{\Delta
t})+\frac{1}{2}\frac{\partial^2p_i}{\partial x^2}(x, t;
n)(-c_i\Delta t-y\sigma_i\sqrt{\Delta t})^2+o(\Delta t)
\right]\rmd y
\]
\[
=\frac{1}{\Delta
t}\int\limits_{-\infty}^\infty\psi(y)\left[\frac{\partial
p_i}{\partial x}(x, t; n)(-c_i\Delta
t)+\frac{1}{2}\frac{\partial^2p_i}{\partial x^2}(x, t;
n)y^2\sigma_i^2\Delta t+o(\Delta t) \right]\rmd y
\]
\[
\to -c_i\frac{\partial p_i}{\partial x}(x, t;
n)+\frac{\sigma_i^2}{2}\frac{\partial^2p_i}{\partial x^2}(x, t;
n),
\]

\noindent so system (\ref{eqDT}) follows from (\ref{eqpipre}) and (\ref{convpi1}). \endproof

It is easy to solve system (\ref{eqDT}).
Let us define functions $q_i=q_i(x, t; n)$. For $n\geq 1$
\begin{eqnarray}\label{fnqeven}
q_0(x, t; 2n)
=\frac{\lambda_0^{n}\lambda_1^{n}}{(c_0-c_1)^{2n}}\cdot\frac{(c_0t-x)^{n-1}(x-c_1t)^n}{(n-1)!n!}\theta(x,
t),
\cr\cr
q_1(x, t; 2n)
=\frac{\lambda_0^{n}\lambda_1^{n}}{(c_0-c_1)^{2n}}\cdot\frac{(c_0t-x)^{n}(x-c_1t)^{n-1}}{n!(n-1)!}\theta(x,
t),
\end{eqnarray}

\noindent and for $n\geq 0$

\begin{eqnarray}\label{fnqodd}
q_0(x, t;
2n+1)=\frac{\lambda_0^{n+1}\lambda_1^{n}}{(c_0-c_1)^{2n+1}}\cdot\frac{(c_0t-x)^{n}(x-c_1t)^n}{(n!)^2}\theta(x,
t),
\cr\cr
q_1(x, t; 2n+1)=\frac{\lambda_0^{n}\lambda_1^{n+1}}{(c_0-c_1)^{2n+1}}\cdot\frac{(c_0t-x)^{n}(x-c_1t)^n}{(n!)^2}\theta(x,
t).
\end{eqnarray}

Here $\theta(x,
t)=\exp\left\{-\frac{\lambda_1}{c_0-c_1}(c_0t-x)-\frac{\lambda_0}{c_0-c_1}(x-c_1t)\right\}
\mathbf{1}_{\{c_1t<x<c_0t\}}$.

The distribution densities $p_i^{(0)}$ of the jump telegraph process
without a diffusion term can be expressed as follows.
Resolving equation (\ref{eqDT}) with $\sigma_0=\sigma_1=0$ we have
\begin{equation}\label{pi0}
p_i^{(0)}(x, t; n)=q_i(x-j_i(n), t; n),
\end{equation}
where $j_i(n)=[(n+1)/2]h_i+[n/2]h_{1-i},\ n=0, 1, \ldots$.
 Equation (\ref{n0}) now means that\\
 $p^{(0)}_0(x, t; 0)=\rme^{-\lambda_0t}\delta(x-c_0t)$,
$p^{(0)}_1(x, t; 0)=\rme^{-\lambda_1t}\delta(x-c_1t)$.

Conditioning on the number of switches we get the probability
density of the jump telegraph process which is described by parameters
$<c_0, \lambda_0, h_0>$ and $<c_1, \lambda_1, h_1>$:
 \begin{equation}\label{a4}
p_{i}^{(0)}(x, t)=\sum\limits_{n=0}^\infty p_{i}^{(0)}(x, t; n).
\end{equation}

\begin{rem}
Formula (\ref{a4})  in particular case $B=h_0+h_1=0$
becomes
\[
p_{i}^{(0)}(x, t)=\rme^{-\lambda_i t}\cdot\delta(x-c_i
t)\]\[+\frac{\theta(x, t)}{c_0-c_1}\left[\lambda_i\exp\left(\frac{\lambda_0-\lambda_1}{c_0-c_1}h_i\right)
I_0\left(2\frac{\sqrt{\lambda_0\lambda_1(c_0t-x+h_i)(x-h_i-c_1t)}}{c_0-c_1}\right)
\right.
\]
\[
\left.+\sqrt{\lambda_0\lambda_1}\left(\frac{x-c_1t}{c_0t-x}\right)^{\frac{1}{2}-i}
I_1\left(2\frac{\sqrt{\lambda_0\lambda_1(c_0t-x)(x-c_1t)}}{c_0-c_1}\right)\right],
\]
where $I_0(z)=\sum_{n=0}^\infty\frac{(z/2)^{2n}}{(n!)^2}$ and
$I_1(z)=I_0'(z)$ are modified Bessel functions. Compare with
Beghin et al (2001).
\end{rem}

We apply previous results to obtain the distributions of times
which the process $\epi$ spends in the certain state.

Let $T_i=T_i(t)=\int_0^t\1_{\{\varepsilon_i(\tau)=0\}}\rmd \tau,\ i=0, 1$ be
the total time between $0$ and $t$ spending by the process
$\varepsilon_i$ in the state $0$ starting form the state $i$.

If we consider a standard telegraph processes with velocities
$c_0=1, c_1=-1$, $\overline{\mathcal{T}_0}(t)=\int_0^t(-1)^{N_0(\tau)}\rmd\tau$ and
$\overline{\mathcal{T}_1}(t)=-\int_0^t(-1)^{N_1(\tau)}\rmd\tau$, then

\begin{equation}\label{XT}
\overline{\mathcal{T}_0}(t)=T_0-(t-T_0)=2T_0-t~~{\rm and}~~\overline{\mathcal{T}_1}(t)=2T_1-t.
\end{equation}

Let  $f_i(\tau, t; n), 0\leq \tau\leq t$ denote the density of $T_i$:
 for all measurable $\Upsilon\subset[0, t]$

\begin{equation}\label{fin}
\int_\Upsilon f_i(\tau, t; n)\rmd \tau=\mathbb{P}_i\{T_i\in\Upsilon,\
N_i(t)=n\}
\end{equation}

\noindent Applying (\ref{XT}) we can notice that

\begin{equation}\label{pf}
f_0(\tau, t; n)=2\bar p_0(2\tau-t, t; n), \qquad f_1(\tau, t; n)=2\bar
p_1(2\tau-t, t; n),
\end{equation}

\noindent where $\bar p_0$ and $\bar p_1$ are the densities of the
standard telegraph process $\overline{\mathcal{T}_0}$ and $\overline{\mathcal{T}_1}$.
Functions   $\bar p_0$ and $\bar p_1$ are  defined in
(\ref{fnqeven})-(\ref{pi0}) with $c_0=1$, $c_1=-1$ and $h_0=h_1=0$.

Using formulae for densities $\bar p_i$, which are obtained in
(\ref{fnqeven})-(\ref{pi0}), from (\ref{pf}) we have
$$
f_0(\tau, t; 0)=\rme^{-\lambda_0t}\delta(\tau-t),\quad f_1(\tau, t;
0)=\rme^{-\lambda_1t}\delta(\tau).
$$
For $n\geq 1$
\begin{equation}\label{f0even}
f_0(\tau, t;
2n)=\lambda_0^n\lambda_1^n\frac{(t-\tau)^{n-1}\tau^n}{(n-1)!n!}\rme^{-\lambda_0\tau-\lambda_1(t-\tau)
}\1_{\{0\leq \tau\leq t\}},
\end{equation}
\begin{equation}\label{f1even}
f_1(\tau, t;
2n)=\lambda_0^n\lambda_1^n\frac{(t-\tau)^{n}\tau^{n-1}}{(n-1)!n!}\rme^{-\lambda_0\tau-\lambda_1(t-\tau)
}\1_{\{0\leq \tau\leq t\}},
\end{equation}
and for $n\geq 0$
\begin{equation}\label{f0odd}
f_0(\tau, t;
2n+1)=\lambda_0^{n+1}\lambda_1^n\frac{(t-\tau)^{n}\tau^n}{(n!)^2}\rme^{-\lambda_0\tau-\lambda_1(t-\tau)
}\1_{\{0\leq \tau\leq t\}},
\end{equation}
\begin{equation}\label{f1odd}
f_1(\tau, t;
2n+1)=\lambda_0^n\lambda_1^{n+1}\frac{(t-\tau)^{n}\tau^n}{(n!)^2}\rme^{-\lambda_0\tau-\lambda_1(t-\tau)
}\1_{\{0\leq \tau\leq t\}}.
\end{equation}

Summarizing we have the following expressions for the densities
$f_i(\tau, t)$ of the spending time of the the process $\epi(\tau), 0\leq
\tau\leq t$ in state $0$:

\[
 f_0(\tau,
t)=\rme^{-\lambda_0t}\delta(\tau-t)+\rme^{-\lambda_0\tau-\lambda_1(t-\tau)
}\left[\lambda_0I_0(2\sqrt{\lambda_0\lambda_1\tau(t-\tau)})\right.
\]
\begin{equation}\label{f0}
\left.+\sqrt{\lambda_0\lambda_1}\sqrt{\frac{\tau}{t-\tau}}I_1(2\sqrt{\lambda_0\lambda_1\tau(t-\tau)})\right]\1_{\{0\leq \tau\leq t\}},
\end{equation}

\[
f_1(\tau,
t)=\rme^{-\lambda_1t}\delta(\tau)+\rme^{-\lambda_0\tau-\lambda_1(t-\tau)
}\left[\lambda_1I_0(2\sqrt{\lambda_0\lambda_1\tau(t-\tau)})\right.
\]
\begin{equation}\label{f1}
\left.+\sqrt{\lambda_0\lambda_1}\sqrt{\frac{t-\tau}{\tau}}I_1(2\sqrt{\lambda_0\lambda_1\tau(t-\tau)})\right]\1_{\{0\leq \tau\leq t\}}.
\end{equation}

In terms of $f_i(\tau, t)$ it is possible to express
the distribution of the telegraph-diffusion process.
If $T_i(t)=\int\limits_0^t\1_{\{\epi(\tau)=0\}}\rmd\tau$, then
$\mathcal{T}_i(t)=c_0T_i(t)+c_1(t-T_i(t))$ and $\mathcal{D}_i(t)\stackrel{d}{=}\sigma_0w(T_i(t))+\sigma_1w'(t-T_i(t))$,
where $w$ and $w'$ are independent.

Let $a_\tau=c_0\tau+c_1(t-\tau)$ and $\Sigma_\tau^2=\sigma_0^2\tau+\sigma_1^2(t-\tau)$.
The distribution densities of telegraph-diffusion process
$\mathcal{T}_i(t)+\mathcal{D}_i(t),\ t\geq 0$ can be expressed as follows:
\[
p_i(x, t)=\frac{1}{\sqrt{2\pi}}\int_0^t\frac{f_i(\tau, t)}
{\Sigma_\tau}\exp\left\{-\frac{1}{2\Sigma_\tau^2}(x-a_\tau)^2\right\}\rmd\tau.
\]

Next, we describe in this framework martingales and martingale
measures. The following theorem could be considered as a version of the
Doob-Meyer decomposition for telegraph-diffusion processes with
alternating intensities.

\begin{theo}\label{prop1}
Jump telegraph-diffusion process $\mathcal{T}_i+\mathcal{J}_i+\mathcal{D}_i, i=0, 1$
is a martingale if and only if
$c_0=-\lambda_0h_0$ and $c_1=-\lambda_1h_1$.
\end{theo}\label{Thm1}

\proof The processes $\sigma_{\epi(t)}, t\geq 0, i=0, 1$ are $\mathfrak{F}_t$-measurable. Hence the processes
$\mathcal{D}_i=\mathcal{D}_i(t)=\int_0^t\sigma_{\epi(\tau)}\rmd w(\tau)$, $t\geq 0$, $i=0,
1$ are $\mathfrak{F}_t$-martingales. Now, the result follows from
Theorem 2.1 of Ratanov (2007).
\endproof

Let $h_0, h_1>-1$. Denote
\begin{equation}\label{kappa}
\kappa_i(t)=\prod\limits_{k=1}^{N_i(t)}(1+h_{\epi(\tau_k-)}).
\end{equation}
\begin{cor}\label{cor11}
The process $\exp\{\mathcal{T}_i(t)+\mathcal{D}_i(t)\}\kappa_i(t)$ is a martingale if
and only if $c_i+\sigma_i^2/2=-\lambda_ih_i,\ i=0, 1.$
\end{cor}

\proof It is sufficient to notice that
$\exp\{\mathcal{T}_i(t)+\mathcal{D}_i(t)\}\kappa_i(t)
=\mathcal{E}_t(\mathcal{T}_i+\mathcal{J}_i+\mathcal{D}_i+1/2\int_0^t\sigma_{\epi(\tau)}^2\rmd\tau)$,
where $\mathcal{E}_t(\cdot)$ denote a stochastic exponential (see Protter (1990)).
The corollary follows from Theorem \ref{prop1}.
\endproof

Now we study the properties of jump telegraph-diffusion processes under a change of
measure. Let $\mathcal{T}^*_i$, $i=0, 1$ be the telegraph processes with
states $<c_0^*, \lambda_0>$ and $<c_1^*, \lambda_1>$, and
$J^*_i=-\sum\limits_{j=1}^{N_i(t)}c^*_{\epi(\tau_j-)}/\lambda_{\ep_i(\tau_j-)}$,
$i=0, 1$ be  the jump processes with jump values
$h_i^*=-c^*_{i}/\lambda_{i}>-1$, which let the sum
$\mathcal{T}_i^*+\mathcal{J}_i^*$ to be a martingale. Let
$\mathcal{D}_i^*=\int\limits_0^t\sigma^*_{\epi(\tau)}\rmd w(\tau)$ be the
diffusion with alternating diffusion coefficients $\sigma_i^*, i=0,
1$. Consider a probability measure $\P_i^*$ with a local density
with respect to $\P_i$:

\begin{equation}\label{Zi}
Z_i(t)=\frac{\P_i^*}{\P_i}|_{t}=\mathcal{E}_t(\mathcal{T}_i^*+\mathcal{J}_i^*+\mathcal{D}_i^*)
=\exp\left({\mathcal{T}_i^*(t)+\mathcal{D}_i^*(t)-\frac{1}{2}\int_0^t{(\sigma^*_{\epi(s)})}^2\rmd
s}\right)\kappa_{i}^*(t),
\end{equation}

\noindent where $\kappa_{i}^*(t)$ is defined in (\ref{kappa}) with
$h_i^*$ instead of $h_i$.

\begin{theo}[Girsanov theorem]\label{Girsanov theorem}
Under the probability measure $\P_i^*$

1) process $\tilde
w(t):=w(t)-\int\limits_0^t\sigma_{\ep_i(\tau)}^*\rmd \tau$ is a
standard Brownian motion;

2) counting Poisson process $N_i(t)$ has intensities
$\lambda_i^*:=\lambda_i(1+h_i^*)=\lambda_i-c_i^*$.
\end{theo}

\proof Let $U_i(t):=\exp\{z\tilde
w(t)\}=\exp\{z(w(t)-\int_0^t\sigma_{\ep_i(\tau)}^*\rmd \tau)\}$.
For 1) it is sufficient to show that for any $t_1<t$

\[
\E_i\{Z_i(t)U_i(t)\ |\
\mathcal{F}_{t_1}\}=\rme^{z^2(t-t_1)/2}Z_i(t_1)U_i(t_1).
\]

\noindent We prove it for $t_1=0$ (see (\ref{XX'})).

Notice that

\[
Z_i(t)U_i(t)=\exp\left\{\mathcal{T}_i^*(t)+\mathcal{D}_i^*(t)-
\frac{1}{2}\int\limits_0^t(\sigma_{\ep_i(\tau)}^*)^2\rmd\tau+zw(t)
-z\int\limits_0^t\sigma_{\ep_i(\tau)}^*\rmd\tau\right\}\kappa_i^*(t)
\]
\[
=\exp\left\{\int\limits_0^t
\left(c_{\epi(\tau)}-\frac{1}{2}{\sigma_{\ep_i(\tau)}^*}^2-z\sigma_{\ep_i(\tau)}^*\right)\rmd\tau
+\int\limits_0^t(\sigma_{\ep_i(\tau)}^*+z)\rmd w(\tau)\right\}\kappa_i^*(t)
\]
\[
=\mathcal{E}_t\left(\mathcal{T}_i^*+\mathcal{D}_i^*+\mathcal{J}_i^*+zw\right)\exp(z^2t/2).
\]

\noindent Thus $\mathbb{E}_i(Z_i(t)U_i(t))=\exp(z^2t/2)$.

To prove the second part of the theorem we denote
$\pi_i^*(t; n)=\mathbb{P}_i^*\{N_i(t)=n\}
=\mathbb{E}_i(Z_i(t)\1_{\{N_i(t)=n\}})
=\kappa_{i}^*(n)\int_{-\infty}^\infty\rme^xp_i^*(x, t; n)\rmd x$,
where $\kappa_{i}^*(n)=\prod\limits_{k=1}^{n}(1+h_{\epi(\tau_k-)})$, and
$p_i^*=p_i^*(x, t; n)$ are (generalized) probability densities of
telegraph-diffusion process
$X_i^*(t)+D_i^*(t)-\int_0^t{(\sigma^*_{\ep_i(\tau)})}^2\rmd
\tau/2$. Notice that functions $p_i^*(x, t; n)$ satisfy system
(\ref{eqDT}) with $c_i^*-(\sigma_i^*)^2/2$ and $\sigma_i^*$
instead of $c_i$ and $\sigma_i$ respectively. Therefore

\[
\frac{\rmd\pi_i^*(t; n)}{\rmd t}=(c_i^*-\lambda_i)\pi_i^*(t; n)+\lambda_i(1+h_i^*)\pi_{1-i}^*(t; n-1).
\]

\noindent Next notice that
$\lambda_i-c_i^*=\lambda_i+\lambda_ih_i^*:=\lambda_i^*$ and, thus

\[
\frac{\rmd\pi_i^*(t; n)}{\rmd t}=-\lambda_i^*\pi_i^*(t; n)+\lambda_i^*\pi_{1-i}^*(t; n-1).
\]

The second part of the theorem now follows from (\ref{eqpi}). \endproof

\section{Jump telegraph-diffusion model}\label{secmod}

 Let $\epi=\epi(t)=0, 1,\ t\geq 0$ be a
Markov switching process defined in Section \ref{sec1} which
indicates the current market state.

Consider $\mathcal{T}_i$, $\mathcal{J}_i$ and $\mathcal{D}_i$,
which are defined in (\ref{tjd}). Assume that $h_0, h_1>-1$.
First, we define the market with one risky asset. Assume that the
price of the risky asset which initially is at the state $i$,
follows the equation

\begin{equation}\label{eq-stock}\nonumber
\rmd S(t)=S(t-)\rmd(\mathcal{T}_i(t)+\mathcal{J}_i(t)+\mathcal{D}_i(t)),\quad i=0, 1.
\end{equation}

As it is observed  in Section \ref{sec1},

\begin{equation}\label{model}
S(t)=S_0\mathcal{E}_t(\mathcal{T}_i+\mathcal{J}_i+\mathcal{D}_i)
=S_0\exp\left(\mathcal{T}_i(t)+\mathcal{D}_i(t)-\frac{1}{2}\int_0^t\sigma_{\epi(\tau)}^2\rmd\tau\right)\kappa_i(t).
\end{equation}

Let $r_i, r_i\geq 0$ is the interest rate of the market which is
at the state $i,\ i=0, 1$. Let us consider the geometric telegraph
process of the form

\begin{equation}\label{eqB}
B(t)=\exp\left\{\mathcal{Y}_i(t)\right\},\qquad \mathcal{Y}_i(t)=\int\limits_0^tr_{\epi(\tau)}\rmd\tau.
\end{equation}

\noindent as a numeraire.

The model (\ref{model})-(\ref{eqB}) is incomplete. Due to simplicity of this  model
 the set $\mathcal{M}$ of equivalent risk-neutral measures can be described in detail.

Let us define an equivalent measure $\P^*_i$ by means of the density $Z_i(t)$ (see (\ref{Zi})) with
$c_i^*$, $h_i^*=-c_i^*/\lambda_i>-1$ and with arbitrary $\sigma_i^*$. Due to Theorem \ref{Girsanov theorem}
$c_i^*=\lambda_i-\lambda_i^*<\lambda_i, i=0,1.$

Let $\theta_0, \theta_1>0$.
We denote $c_0^*=\lambda_0-\theta_0$, $c_1^*=\lambda_1-\theta_1$,
$h_0^*=-1+\theta_0/\lambda_0$, $h_1^*=-1+\theta_1/\lambda_1$, and we take
arbitrary $\sigma^*_0$, $\sigma^*_1$.
Due to Theorem \ref{Girsanov theorem}, under the measure $\P_i^*$ the driving Poisson process $N_i(t)$
has intensities $\lambda_i^*=\theta_i, i=0, 1$. The equivalent risk-neutral measures
for the model (\ref{model})-(\ref{eqB}) depend on two positive parameters $\theta_0$ and $\theta_1$.

 \begin{theo}\label{th31}
 Let probability measure $\P_i^*$ be defined by
 means of the density $Z_i(t), t\geq 0$. Let $\sigma_0\neq 0$ and $\sigma_1\neq 0$.
 The process $B(t)^{-1}S(t)$ is a
 $\P_i^*$-martingale if and only if the measure $\P_i^*$ is defined
 by parameters $c_0^*=\lambda_0-\theta_0$, $c_1^*=\lambda_1-\theta_1$,
$h_0^*=-1+\theta_0/\lambda_0$, $h_1^*=-1+\theta_1/\lambda_1$ and
 $\sigma_0^*$ and $\sigma_1^*$ which are as follows:
 $\sigma_0^*=(r_0-c_0-h_0\theta_0)/\sigma_0$ and
 $\sigma_1^*=(r_1-c_1-h_1\theta_1)/\sigma_1$, $\theta_0, \theta_1>0$.
 \end{theo}

\proof Indeed,

$$Z_i(t)B(t)^{-1}S(t)=S_0\exp\{{Y}_i(t)\}\tilde\kappa_i(t),$$

\noindent where

$${Y}_i(t)=\mathcal{T}_i(t)+\mathcal{T}_i^*(t)+\mathcal{D}_i(t)+\mathcal{D}_i^*(t)-
\frac{1}{2}\int\limits_0^t({\sigma_{\epi(\tau)}}^2
+{\sigma^*_{\epi(\tau)}}^2)\rmd\tau-\mathcal{Y}_i(t)$$

\noindent and $\tilde\kappa_i(t)$ is defined as in (\ref{kappa})
with $\tilde h_i$  instead of $h_i$. Here $\tilde h_i$ satisfies the equation
\[
1+\tilde h_i=(1+h_i^*)(1+h_i),\ i=0, 1.
\]
Thus $\tilde h_i=h_i+h_i^*+h_ih_i^*
=h_i+(-1+\theta_i/\lambda_i)+h_i(-1+\theta_i/\lambda_i)
=\theta_i(1+h_i)/\lambda_i-1,\ i=0, 1$.
Using Corollary \ref{cor11} we see that $Z_i(t)B(t)^{-1}S(t)$ is
the $\P_i$-martingale, if and only if

\[
\begin{cases}
c_0+c_0^*-r_0+\sigma_0\sigma_0^*=-\lambda_0\tilde h_0\\
c_1+c_1^*-r_1+\sigma_1\sigma_1^*=-\lambda_1\tilde h_1
\end{cases}.
\]

Note that $c_i^*=\lambda_i-\theta_i$ and $\lambda_i\tilde h_i=\theta_i(1+h_i)-\lambda_i$, so

\[
\begin{cases}
c_0+(\lambda_0-\theta_0)-r_0+\sigma_0\sigma_0^*=-\theta_0(1+h_0)+\lambda_0\\
c_1+(\lambda_1-\theta_1)-r_1+\sigma_1\sigma_1^*=-\theta_1(1+h_1)+\lambda_1
\end{cases},
\]
and then
\begin{equation}\label{crsigma}
\begin{cases}
c_0-r_0+\sigma_0\sigma_0^*=-\theta_0 h_0\\
c_1-r_1+\sigma_1\sigma_1^*=-\theta_1 h_1
\end{cases}.
\end{equation}

Therefore $\sigma_i^*=(r_i-c_i-h_i\theta_i)/\sigma_i,\ i=0, 1.$
\endproof

\begin{rem}
The case of $\sigma_0=\sigma_1=0$ is called jump-telegraph model, and it is complete.
In this case the martingale measure is defined by $c_i^{*}=\lambda_i-\lambda_i^{*}$ and
$\lambda_i^{*}=\frac{r_i-c_i}{h_i}$ as the new intensities of switchings.
See  Ratanov (2007) for details.

The Black-Scholes model respects to $h_0=h_1=0,\ \sigma_0=\sigma_1:=\sigma,\ c_0=c_1:=c,\ r_0=r_1=r$.
In this case system (\ref{crsigma}) has the unique solution $\sigma_0^*=\sigma_1^*=\sigma^*=\frac{r-c}{\sigma}$.
It means that the martingale measure is unique. Due to Girsanov theorem \ref{Girsanov theorem} the process $w(t)-\sigma^*t$
is Brownian motion under the new measure, which repeats the classic result.
\end{rem}

To complete the model we add a new asset. Consider the market
of two risky assets which are driven by common Brownian motion $w$ and
counting Poisson processes $N_i$:

\begin{equation}\label{two1}
\rmd
S^{(1)}(t)
=S^{(1)}(t-)\rmd(\mathcal{T}^{(1)}_i(t)+\mathcal{J}^{(1)}_i(t)+\mathcal{D}^{(1)}_i(t)),
\end{equation}
\begin{equation}\label{two2}
\rmd
S^{(2)}(t)
=S^{(2)}(t-)\rmd(\mathcal{T}^{(2)}_i(t)+\mathcal{J}^{(2)}_i(t)+\mathcal{D}^{(2)}_i(t)).
\end{equation}

\noindent As usual,  $i=0, 1$ denotes the initial market state.

Denote

$$
\Delta_0^{(h)}=\left|\begin{array}{cc}\sigma_0^{(1)}&h_0^{(1)}\\
                                 \sigma_0^{(2)}&h_0^{(2)}\end{array}\right|
                                 =\sigma_0^{(1)}h_0^{(2)}-\sigma_0^{(2)}h_0^{(1)},\quad
\Delta_1^{(h)}=\left|\begin{array}{cc}\sigma_1^{(1)}&h_1^{(1)}\\
                                 \sigma_1^{(2)}&h_1^{(2)}\end{array}\right|
                                 =\sigma_1^{(1)}h_1^{(2)}-\sigma_1^{(2)}h_1^{(1)},
$$

\noindent and

$$
\Delta_0^{(r-c)}=\left|\begin{array}{cc}\sigma_0^{(1)}&r_0-c_0^{(1)}\\
                                 \sigma_0^{(2)}&r_0-c_0^{(2)}\end{array}\right|
                                 =\sigma_0^{(1)}(r_0-c_0^{(2)})-\sigma_0^{(2)}(r_0-c_0^{(1)}),$$$$
\Delta_1^{(r-c)}=\left|\begin{array}{cc}\sigma_1^{(1)}&r_1-c_1^{(1)}\\
                                 \sigma_1^{(2)}&r_1-c_1^{(2)}\end{array}\right|
                                 =\sigma_1^{(1)}(r_1-c_1^{(2)})-\sigma_1^{(2)}(r_1-c_1^{(1)}).
$$

Let $\Delta_0^{(h)}\neq 0,\ \Delta_1^{(h)}\neq 0$. We  assume that

\begin{equation}\label{lambda*}
\lambda_i^*:=\frac{\Delta_i^{(r-c)}}{\Delta_i^{(h)}}>0.
\end{equation}

\begin{theo}\label{MM}
Both processes $B(t)^{-1}S^{(m)}(t), t\geq 0, m=1, 2$ are
$\P^*_i$-martingales if and only if the measure $\P^*_i$ is
defined by (\ref{Zi}) with the following parameters: 

\begin{equation}\label{sigma*}
\sigma_0^*
=\frac{(r_0-c_0^{(1)})h_0^{(2)}-(r_0-c_0^{(2)})h_0^{(1)}}{\Delta_0^{(h)}},\qquad
\sigma_1^*
=\frac{(r_1-c_1^{(1)})h_1^{(2)}-(r_1-c_1^{(2)})h_1^{(1)}}{\Delta_1^{(h)}},
\end{equation}

\begin{equation}\label{c*}
c_0^*
=\lambda_0-\frac{\Delta_0^{(r-c)}}{\Delta_0^{(h)}},\qquad
c_1^*
=\lambda_1-\frac{\Delta_1^{(r-c)}}{\Delta_1^{(h)}}
\end{equation}

\noindent and

$$
 h_0^*=-c_0^*/\lambda_0,\qquad h_1^*=-c_1^*/\lambda_1.
$$

Under the measure $\P_i^*$ the rate of leaving the
state $i$ equals to $\lambda_i^*$ defined in (\ref{lambda*}).
\end{theo}

\proof First notice

$$
Z_i(t)B(t)^{-1}S^{(m)}(t)
=S^{(m)}(0)\mathcal{E}_t\exp(\mathcal{T}_i^*+\mathcal{J}_i^*+\mathcal{D}_i^*)
\exp(-Y_i(t))\mathcal{E}_t(\mathcal{T}_i^{(m)}+\mathcal{J}_i^{(m)}+\mathcal{D}_i^{(m)})
$$
$$
=\exp\left(\mathcal{T}_i^*(t)+\mathcal{D}_i^*(t)-\frac{1}{2}\int\limits_0^t{\sigma^*_{\epi(\tau)}}^2\rmd\tau\right)\kappa_i^*(t)
$$$$\times\exp\left(\mathcal{T}_i^{(m)}(t)+\mathcal{D}_i^{(m)}(t)-Y_i(t)-\frac{1}{2}\int\limits_0^t{\sigma_{\epi(\tau)}^{(m)}}^2\rmd\tau\right)\kappa_i^{(m)}(t)
$$
$$
=\mathcal{E}_t\left(\mathcal{T}_i^{(m)}+\mathcal{T}_i^*+\mathcal{D}_i^{(m)}+\mathcal{D}_i^*
-Y_i+\int\limits_0^t\sigma_{\epi(\tau)}^{(m)}\sigma_{\epi(\tau)}^*\rmd\tau\right)\kappa_i^{(m)}(t)\kappa_i^*(t).
$$

Thus $Z_i(t)B(t)^{-1}S^{(m)}(t)$ is a martingale if and only if (Theorem \ref{prop1})

\begin{equation}\label{eqcompl1}
\begin{cases}
c_i^{(1)}+{c_i^*}-r_i+\sigma_i^{(1)}{\sigma_i^*}
=-\lambda_i(h_i^{(1)}+{h_i^*}+h_i^{(1)}{h_i^*})\cr
c_i^{(2)}+{c_i^*}-r_i+\sigma_i^{(2)}{\sigma_i^*}
=-\lambda_i(h_i^{(2)}+{h_i^*}+h_i^{(2)}{h_i^*})
\end{cases}.
\end{equation}

Now using the identities $c_i^*=-\lambda_ih_i^*, i=0, 1$ we simplify the system (\ref{eqcompl1}) to

\begin{equation}\label{eqcompl2}
\begin{cases}
\sigma_i^{(1)}{\sigma_i^*}-h_i^{(1)}c_i^*
=r_i-c_i^{(1)}-\lambda_ih_i^{(1)}\cr
\sigma_i^{(2)}{\sigma_i^*}-h_i^{(2)}c_i^*
=r_i-c_i^{(2)}-\lambda_ih_i^{(2)}
\end{cases}.
\end{equation}

Systems (\ref{eqcompl2}) have the solutions described in (\ref{sigma*})-(\ref{c*}).

Note that as it follows from Girsanov theorem, the intensity parameters under measure $\P_i^*$,
$\lambda_0^*$ and $\lambda_1^*$
are defined in (\ref{lambda*}).
\endproof

\begin{cor} Let $\Delta_0^{(h)}\neq 0,\ \Delta_1^{(h)}\neq 0$ and (\ref{lambda*}) is fulfilled.
If the prices $S_i^{(1)}$ and $S_i^{(2)}$ of both risky assets are are defined in (\ref{two1})-(\ref{two2}) with nonzero
jumps, $h_0^{(m)}\neq 0, h_1^{(m)}\neq 0,\ m=1, 2$, then

$$
\sigma_0^*
=\frac{\alpha_0^{(1)}-\alpha_0^{(2)}}{\beta_0^{(1)}-\beta_0^{(2)}},\qquad \sigma_1^*
=\frac{\alpha_1^{(1)}-\alpha_1^{(2)}}{\beta_1^{(1)}-\beta_1^{(2)}}
$$

\noindent and

$$
c_0^*
=\lambda_0-\frac{\beta_0^{(1)}\alpha_0^{(2)}-\beta_0^{(2)}\alpha_0^{(1)}}{\beta_0^{(1)}-\beta_0^{(2)}},\qquad
c_1^*
=\lambda_1-\frac{\beta_1^{(1)}\alpha_1^{(2)}-\beta_1^{(2)}\alpha_1^{(1)}}{\beta_1^{(1)}-\beta_1^{(2)}},
$$

\noindent where
$$
\alpha_0^{(m)}=\frac{r_0-c_0^{(m)}}{h_0^{(m)}},\qquad \alpha_1^{(m)}=\frac{r_1-c_1^{(m)}}{h_1^{(m)}},\qquad
\beta_0^{(m)}=\frac{\sigma_0^{(m)}}{h_0^{(m)}},\qquad \beta_1^{(m)}=\frac{\sigma_1^{(m)}}{h_1^{(m)}},\qquad m=1, 2.
$$
\end{cor}

\begin{rem}
If $\Delta_0^{(h)}=\Delta_1^{(h)}=0$, then the system (\ref{eqcompl2}) does not have a solution
(if $\Delta_0^{(r-c)}\neq 0$, $\Delta_1^{(r-c)}\neq 0$)
 or it has infinitely many solutions (if $\Delta_0^{(r-c)}=\Delta_1^{(r-c)}=0)$.
 It means arbitrage or incompleteness respectively.

In particular case of the market model without jumps, i. e.  $h_i^{(1)}=h_i^{(2)}=0, i=0, 1$,
the market of two assets is arbitrage-free (respectively, the system (\ref{eqcompl2}) has solutions) if and only if the assets are similar:
$$\frac{r_i-c_i^{(1)}}{\sigma_i^{(1)}}=\frac{r_i-c_i^{(2)}}{\sigma_i^{(2)}}=\sigma_i^{*}, \ i=0, 1.$$

In this case the model is incomplete.
\end{rem}

\begin{rem}
Hidden Markov model with $h_0^{(1)}=h_1^{(1)}=0$
can be completed by adding a security that pays one unit of bond
 at the next time that the Markov chain $\epi(t)$ changes state (see Guo (2001)).
 That change-of-state contract then becomes worthless and a new contract is issued
 that pays at the next change of state, and so on.
Under natural pricing, this completes the model, and $\lambda_i^*=\frac{r_i\lambda_i}{r_i+k_i}$,
where $k_i$ is given, and can be thought as a risk-premium coefficient.

Theorem \ref{th31} presents the unique risk-neutral measure for this completion of the market. It is given by (\ref{Zi}) with
$c_i^*=\lambda_i-\lambda_i^*=\frac{\lambda_ik_i}{r_i+k_i},\
h_i^*=-1+\lambda_i^*/\lambda_i=-\frac{k_i}{r_i+k_i}$ and $\sigma_i^*=(r_i-c_i)/\sigma_i,\ i=0, 1$.

In our framework the stock (without jump component)
\[
S^{(1)}(t)=S^{(1)}(0)\rme^{\mathcal{T}_i(t)+\mathcal{D}_i(t)-\frac{1}{2}\int_0^{t}\sigma_{\epi(\tau)}^{2}\rmd\tau}
\]
can be naturally accompanied with the security
which magnifies its value with the fixed rate at each moment of the change of state:
\[
S^{(2)}(t)=\prod\limits_{k=1}^{N_i(t)}(1+h_{\epi(\tau_k-)}), \qquad h_0, h_1>0.
\]
This security can be considered  as an insurance contract,
different from change-of-state contract proposed in Guo (2001), that
compensates losses provoked by state changes.

By the definition we see that $\Delta_i^{(h)}=\sigma_ih_i,\ \Delta_i^{(r-c)}=\sigma_ir_i$.
Thus Theorem \ref{MM} gives
$\lambda_i^{*}=r_i/h_i,\ \sigma_i^{*}=(r_i-c_i)/\sigma_i,\
c_i^{*}=\lambda_i-r_i/h_i,\ h_i^{*}=-1+r_i/(\lambda_ih_i)$.

In contrast with Guo (2001) the security which completes the market model is perpetual, i. e.
it not becomes worthless at the switching times.
\end{rem}

Assume now that $\Delta_0^{(h)}\neq 0$ and $\Delta_1^{(h)}\neq 0$, and (\ref{lambda*}) is fulfilled.
Therefore the market model can be completed.
Let us present the formula for the price of standard call option.
Let $Z$ be a r.v. with normal distribution $\mathcal{N}(0,
\sigma^2)$. We denote
\begin{equation}\label{BSfunction}
\varphi(x, K,
\sigma):=\mathbb{E}[x\rme^{Z-\sigma^2/2}-K]^+
=xF(\frac{\ln(x/K)+\sigma^2/2}{\sigma})-KF(\frac{\ln(x/K)-\sigma^2/2}{\sigma}),
\end{equation}
where $F(x)$ is the distribution function of standard normal law:
$$F(x)=\frac{1}{\sqrt{2\pi}}\int\limits_{-\infty}^x \rme^{-y^2/2}\rmd y.$$

Let the market contains two risky assets (\ref{two1})-(\ref{two2}). Consider the
standard call option on the first asset with the claim
$\left(S^{(1)}(T)-K\right)^+$.
 Therefore the  call-price is

 \begin{equation}\label{callprice1}
\mathfrak{c}_i=\mathbb{E}_i^*\{B(T)^{-1}(S^{(1)}_i(T)-K)^+\},
 \end{equation}

 \noindent if the market is starting with the state $i$. Here $\mathbb{E}_i^*$ is the expectation with respect
 to the martingale measure $\P_i^*$ which is constructed in Theorem
 \ref{MM}.

By Girsanov theorem \ref{Girsanov theorem} the process
$\tilde w(t)=w(t)-\int_0^t\sigma^*_{\varepsilon_i(\tau)}\rmd \tau$ is
the Brownian motion under new measure $\mathbb{P}_i^*$. Hence
\[
B(T)^{-1}S^{(1)}(T)=S^{(1)}(0)\exp\left\{\mathcal{T}^{(1)}_i(T)+\int_0^T\sigma_{\varepsilon_i(\tau)}^{(1)}\rmd
w(\tau)-\frac{1}{2}\int_0^T{\sigma_{\varepsilon_i(\tau)}^{(1)}}^2\rmd
\tau-\mathcal{Y}_i(T)\right\}\kappa^{(1)}_i(T)
\]
\[
=S^{(1)}(0)\exp\left\{\mathcal{T}^{(1)}_i(T)+\int_0^T\sigma_{\varepsilon_i(\tau)}^{(1)}\rmd
\tilde w(\tau)+\int_0^T\sigma_{\varepsilon_i(\tau)}^{(1)}\sigma_{\varepsilon_i(\tau)}^*\rmd\tau
-\frac{1}{2}\int_0^T{\sigma_{\varepsilon_i(\tau)}^{(1)}}^2\rmd
\tau-\mathcal{Y}_i(T)\right\}\kappa^{(1)}_i(T).
\]
The first equation of (\ref{eqcompl2}) can be transformed to
$c_i^{(1)}-r_i+\sigma_i^{(1)}\sigma_i^*=h_i^{(1)}(c_i^*-\lambda_i)$. From Girsanov theorem
\ref{Girsanov theorem} we have $c_i^{*}-\lambda_i=-\lambda_i^*$. Let us introduce the
 telegraph process $\overline{\mathcal{T}_i^{(1)}}$ independent of $\tilde w$ which is driven
by Poisson process with parameters $\lambda_i^*$ and with the
velocities $\tilde{c_i}=c_i^{(1)}-r_i+\sigma_i^{(1)}\sigma_i^{*}=-\lambda_i^*h_i^{(1)}$, $i=0, 1$.
So the martingale $B(T)^{-1}S^{(1)}(T)$ takes the form
\[
B(T)^{-1}S^{(1)}(T)=S^{(1)}(0)\exp\left\{
\overline{\mathcal{T}^{(1)}_i}(T)+\int_0^T\sigma_{\varepsilon_i(\tau)}^{(1)}\rmd
\tilde w(\tau)-\frac{1}{2}\int_0^T{\sigma_{\varepsilon_i(\tau)}^{(1)}}^2\rmd
\tau\right\}\kappa^{(1)}_i(T)
\]

Again applying the property (\ref{XX'}), from (\ref{callprice1}) we obtain
\begin{equation}\label{calloptionprice}
\mathfrak{c}_i=\int\limits_0^T\sum\limits_{n=0}^\infty f_i(t, T;
n)\varphi(x_i(t, T, n), K\rme^{-r_0t-r_1(T-t)},
\sqrt{\sigma_0^2t+\sigma_1^2(T-t)})\rmd t,\ i=0, 1.
\end{equation}
Here $x_i(t, T,
n)=S^{(1)}(0)\kappa_{i, n}\rme^{\tilde c_0t+\tilde c_1(T-t)}$ and
$$\kappa_{i, 2n}=(1+h_0^{(1)})^n(1+h_1^{(1)})^n,\ i=0, 1,$$
$$\kappa_{0, 2n+1}=(1+h_0^{(1)})^{n+1}(1+h_1^{(1)})^n,\qquad\kappa_{1, 2n+1}
=(1+h_1^{(1)})^{n+1}(1+h_0^{(1)})^n,$$ $$n=0, 1, 2,
\ldots;$$
$f_i(t, T; n)$ are defined in (\ref{f0even})-(\ref{f1odd})
with $\lambda_0^{*}=\Delta_0^{(r-c)}/\Delta_0^{(h)}$, and $\lambda_1^{*}=\Delta_1^{(r-c)}/\Delta_1^{(h)}$
instead of $\lambda_0$ and $\lambda_1$; $\varphi(x, K, \sigma)$ is defined in (\ref{BSfunction}).
Notice that as in jump-telegraph model (see Ratanov (2007)) the option price (\ref{calloptionprice})
does not depend on $\lambda_0$ and $\lambda_1$.

In particular, if $h_{0}^{(1)}=h_1^{(1)}=0$ and, nevertheless, $\Delta_0^{(h)}\neq 0,\ \Delta_1^{(h)}\neq 0$, we can summarize in
(\ref{calloptionprice}) applying (\ref{f0even})-(\ref{f1odd}):
\begin{equation}\label{calloptionprice0}\nonumber
\mathfrak{c}_i=\int_0^Tf_i(t,
T)\varphi(S_0,
K\rme^{-r_0t-r_1(T-t)}, \sqrt{\sigma_0^2t+\sigma_1^2(T-t)})\rmd t,\ i=0,
1,
\end{equation}
where $f_i(t, T)$ are defined in (\ref{f0}) and (\ref{f1}) (cf.
Guo (2001)).


\begin{thebibliography}{99}

\bibitem{Ba}
{\sc B\"{a}uerle, N. and  K\"{o}tter, M.} (2007).
Markov-modulated diffusion risk models.
\emph{Scandinavian Actuarial J.}, Volume 2007, Number 1, 34-52.



\bibitem{Luisa}
{\sc Beghin, L., Nieddu, L. and Orsingher, E.} (2001). Probabilistic
analysis of the telegrapher's process with drift by mean of
relativistic transformations. {\it J. Appl. Math. Stoch. Anal.}
{\bf 14} 11-25.


\bibitem{BS}
{\sc Black, F. and Scholes, M.} (1973). The pricing of options and corporate liabilities.
\emph{Journal of Political Economy} \textbf{81} 637-654.

\bibitem{CoxRoss}
{\sc Cox, J.C. and Ross, S.} (1976).
The valuation of options for alternative
stochastic processes. \emph{J. Financ. Econ.}  \textbf{3} 145–166.

\bibitem{Clark}
{\sc Clark, P.K.} (1973). A Subordinated Stochastic
Process Model with Finite Variance for Speculative Prices.
\emph{Econometrica}, \textbf{41}, No. 1 (Jan., 1973),  135-155.

\bibitem{Ia}
{\sc De Gregorio, A.  and Iacus, S. M.} (2007).
Change point estimation for the telegraph
process observed at discrete time. {\it Working paper}
Dipartimento di Scienze Economiche, Aziendali e Statistiche,
University of Milan.

\bibitem{DiC}
{\sc Di Crescenzo, A. and Pellerey, F.} (2002). On prices'
evolutions based on geometric telegrapher's process. \emph{ Appl.
Stoch. Models Bus. Ind.} {\bf 18}  171-184.


\bibitem{DiM}
{\sc Di Masi, G., Kabanov, Y. and Runggaldier, W.} (1994).
Mean-variance hedging of options on stocks with Markov
volatilities. {\it Theor. Prob. Appl.}  {\bf 39} 211-222.

\bibitem{Elliott}
{\sc Elliott, R. and van der Hoek J.} (1997).
An application of hidden Markov models to asset allocation problems.
\emph{Finance and Stochastics}, \textbf{1} 229-238.

\bibitem{Gold}
{\sc Goldstein, S.} (1951).  On diffusion by discontinuous
movements and on telegraph equation. \emph{Quart. J. Mech. Appl.
Math.} {\bf 4} 129--156.

\bibitem{XG}
{\sc Guo, X. } (2001). Information and option pricings. {\it
Quant. Finance} \textbf{1}  38-44.



\bibitem{Rogers}
{\sc Jobert, A. and Rogers, L.C.G. } (2006). Option pricing with
Markov-modulated dynamics. \emph{SIAM J. Control Optim.} {\bf 6}
2063-2078.

\bibitem{Kac}
{\sc Kac, M.} (1974). A stochastic model related to the telegraph
equation. \emph{Rocky Mountain J. Math.} {\bf 4}  497-509.

\bibitem{KarShr}
{\sc Karatzas, I. and Shreve, S. E. }  (1998).  {\it Methods of
mathematical finance}, vol. \textbf{39} of {\it Applications of
Mathematics}. Springer-Verlag, New York.

\bibitem{Mandelbrot}
{\sc Mandelbrot, B.} (1963).
The Variation of Certain Speculative Prices.
\emph{The Journal of Business}  \textbf{36} No. 4 (Oct., 1963), 394-419.

\bibitem{MT}
{\sc Mandelbrot, B. and  Taylor, H. }(1967). On the Distribution
of Stock Price Differences. \emph{Operations Research} \textbf{15} 1057-1062.

\bibitem{MMPW}
{\sc Masoliver, J., Montero, M., Perell\'o, J.  and Weiss, G.H.} (2006).
The CTRW in finance: Direct and inverse problems. \emph{J. Econ. Behav. Organ.}
\textbf{61} 577-598.


\bibitem{Mazza}
{\sc Mazza, C. and Rulli\`ere, D.} (2004). { A link between wave
governed random motions and ruin processes}. \emph{Insurance:
Mathematics and Economics} {\bf 35}  205-222.

\bibitem{Mer}
{\sc Merton, R. C.} (1973). Theory of rational option pricing, \emph{Bell Journal of Economics and
Management Science} \textbf{4} 141-183.

\bibitem{Merton}
{\sc Merton, R. C.} (1976).
Option pricing when underlying stock returns are
discontinuous. {\it J. of Financial Economics} {\bf 3} 125-144.

\bibitem{Montero}
{\sc Montero, M.} (2008). Renewal equations for option pricing.
Submitted in \emph{Europ. Phys. J.}

\bibitem{NV}
{\sc Nicolato, E and Venardos, E.} (2003). Option pricing in stochastic
volatility models of the
Ornstein-Uhlenbeck type. \emph{Math. Finance} \textbf{13} 445-466.

\bibitem{Protter}
{\sc Protter, P.} (1990). {\it Stochastic Integration and Differential Equations. A New Approach}.
\textbf{21} of {\it Applications
of Mathematics}, Springer, Berlin.



\bibitem{QF}
{\sc Ratanov, N.} (2007a). {A jump telegraph model for option
pricing}. {\it Quant. Finance} \textbf{7}  575-583.

\bibitem{PAMM}
{\sc Ratanov, N.} (2007b).
An option pricing model based on jump telegraph processes.
\emph{PAMM} \textbf{7}, Issue 1, 2080009-2080010.

\bibitem{Kobayashi}
{\sc Ren, Q. and Kobayashi, H.} (1998).
Diffusion Approximation Modeling for Markov
Modulated Bursty Traffic and Its Applications
to Bandwidth Allocation in ATM Networks.
\emph{IEEE J. on Selected Areas in Comm.} \textbf{16}, No. 5, 679-691.

\bibitem{Zacks}
{\sc Zacks, S.} (2004). {Generalized integrated telegraph
processes and the distribution of related stopping times},
\emph{J. Appl. Prob.} {\bf 41}  497-507.
\end{thebibliography}
\end{document}